\documentclass[%
 reprint,
 amsmath,amssymb,
 aps,
]{revtex4-2}

\usepackage{graphicx}
\usepackage{times}
\usepackage{dcolumn}
\usepackage{bm}

\begin{document}

\preprint{APS/123-QED}

\title{Influence of polydispersity on the relaxation mechanisms of glassy liquids}

\author{Ilian Pihlajamaa}
 \affiliation{Soft Matter and Biological Physics, Department of Applied Physics, Eindhoven University of Technology,
P.O. Box 513, 5600 MB Eindhoven, Netherlands}
\author{Corentin C.L. Laudicina}
 \affiliation{Soft Matter and Biological Physics, Department of Applied Physics, Eindhoven University of Technology,
P.O. Box 513, 5600 MB Eindhoven, Netherlands}
\author{Liesbeth M.C. Janssen}
 \affiliation{Soft Matter and Biological Physics, Department of Applied Physics, Eindhoven University of Technology,
P.O. Box 513, 5600 MB Eindhoven, Netherlands}

\date{\today}

\begin{abstract}
State-of-the-art techniques for simulating deeply supercooled liquids require a high degree of size polydispersity to be effective. While these techniques have enabled great insight into the microscopic dynamics near the glass transition, the effect of the large polydispersity on the dynamics has remained largely unstudied. Here we show that a particle's size not only has a strong correlation with its mobility, but we also observe that, as the mode-coupling temperature is crossed and the system becomes more deeply supercooled, a dynamic separation between small mobile and larger quiescent particles emerges at timescales corresponding to cage escape. Our results suggest that the cage escape of this population of mobile particles facilitates the later structural relaxation of the quiescent particles. In the deep supercooled regime, we show that particles of different sizes display varying degrees of breakdown of the Stokes-Einstein relation and have different activation energy barriers. Overall, this indicates that it is important to account for particle-size effects when generalizing results to other glass-forming systems.
\end{abstract}

\maketitle


\section*{Introduction} 

Recent advances in simulation techniques for the study of polydisperse supercooled liquids have led to significant breakthroughs in our understanding of the dynamics near the glass transition, enabling the equilibration of deeply supercooled states that are unreachable by standard techniques \cite{guiselin2022microscopic,scalliet2022thirty, berthier2017configurational, ozawa2017does, berthier2023modern}. Specifically, Swap Monte Carlo (SMC) is able to efficiently sample the phase space of a system through Monte Carlo moves that swap the diameters of two randomly chosen particles. This allows the system to circumvent large energy barriers in the free energy landscape by using particles' sizes as additional degrees of freedom \cite{grigera2001fast, brumer2004numerical, ninarello2017models, berthier2016equilibrium, berthier2019efficient}. Even though the dynamics produced by this technique are clearly nonphysical, detailed balance is obeyed, ensuring that states are visited in accordance to their \textit{a priori} probability. This makes the method suitable for measurements of static equilibrium properties, as well as for the preparation of well-equilibrated, deeply supercooled configurations from which conventional dynamical simulations can be started.

Before the advent of SMC, benchmark model systems to study supercooled liquids would typically include a minimal number of particle species sufficient to prevent crystallization, with a volume ratio between the largest and smallest species of less than 2 ($\approx $5\% polydispersity) \cite{kob1995testing}.
While such systems can be equilibrated with SMC (after minor modifications \cite{parmar2020stable, parmar2020ultrastable}), swap moves are most effective if the degree of polydispersity is increased. This is the cause of the recent rise in popularity of a continuously polydisperse mixture with a maximal volume ratio of over 10 (polydispersity of 23\%) \cite{ninarello2017models}. Ninarello \textit{et al.}\ have shown that this mixture can be equilibrated under unprecedentedly deep supercooling conditions and remains stable far below the mode-coupling temperature $T_\mathrm{mct}$ \cite{ninarello2017models}, with $T_\mathrm{mct}$ defined as the dynamic crossover point between power-law and Arrhenius scaling of the structural relaxation time with temperature. Below this crossover temperature, the dynamics of the system become increasingly cooperative in the sense that initial reorganizations enable further local relaxations, supporting the picture of dynamic facilitation \cite{scalliet2022thirty, garrahan2002geometrical, chandler2010dynamics, chacko2021elastoplasticity}. While the significance of these results should not be understated as they have led to profound insights in the dynamics of deeply supercooled liquids \cite{berthier2023modern}, it remains unknown  to what extent the high degree of polydispersity of this mixture underpins its dynamical behavior. Consequently, it is also unclear to what extent the findings are representative of other, less polydisperse glass-formers.

Already in less polydisperse mixtures, the introduction of different particle species inherently complicates the dynamics of glass-forming systems \cite{baranau2020relaxation, coslovich2018local, schope2007effect}. Within mildly polydisperse supercooled systems, particles of different sizes are known to have different diffusivities \cite{murarka2003diffusion, heckendorf2017size, behera2017effects, puertas2004dynamical, laurati2018different,higler2019diffusion}, and smaller particles can initiate hopping behavior at slightly higher temperatures than larger ones \cite{flenner2005relaxation}. If the degree of polydispersity is high, there even exist states where small and large particles vitrify at different temperatures \cite{voigtmann2011multiple, vaibhav2022finite}. Despite these results, the precise role of polydispersity in hallmark features of glass-forming liquids such as the dynamic crossover, dynamic heterogeneity, and Stokes-Einstein violation \cite{berthier2011theoretical,li2016influence, abraham2008energy} remains largely unexplored.

This issue is especially relevant in the context of new SMC studies,  
since there is evidence that for degrees of 
polydispersity exceeding 10-12\%, the nature of structural relaxation in supercooled liquids is altered \cite{zaccarelli2015polydispersity}. In such systems, not only the degree of polydispersity, but also the shape of the particle size distribution seems to affect whether or not a system can be thermalized, and the degree to which the mobilities of small and large particles decouple. It is even argued that whether or not the power-law scaling of the relaxation time survives up to a glass transition might depend on the particle size distribution \cite{zaccarelli2015polydispersity, van2010comment, brambilla2009probing}. As the polydispersity in the mixture popularized by SMC greatly exceeds 12\%, it is unknown to what extent the above observations persist and how they relate to the microscopic relaxation mechanisms. Understanding the effects of polydispersity in these mixtures may carry important implications for the interpretation of recent advances in the field of the glass transition, such as the observation of an excess wing in the relaxation spectra and the recent evidence for the picture of dynamic facilitation. While these are clearly important questions, thus far no effort has gone into addressing them.

Here we investigate, for the most deeply supercooled simulation model to date \cite{ninarello2017models, scalliet2022thirty}, how the inherently large polydispersity of this mixture affects its supercooled dynamics across the mode-coupling crossover temperature. Specifically, we resolve the dynamics for different particle sizes and find that near $T_\mathrm{mct}$ a dynamical separation emerges between small mobile and larger trapped particles. Not only does the relaxation of these smaller particles precede that of the larger particles, suggesting that the small particles facilitate the relaxation of the larger ones, but we also find that the main relaxation mechanism itself is fundamentally different between the sub-populations.
Our work highlights that great care must be taken when general conclusions are drawn about fundamental glass physics from particle-size-agnostic quantities in the analysis of highly polydisperse glassy mixtures.

\begin{figure}
    \centering
    \includegraphics[width=\linewidth]{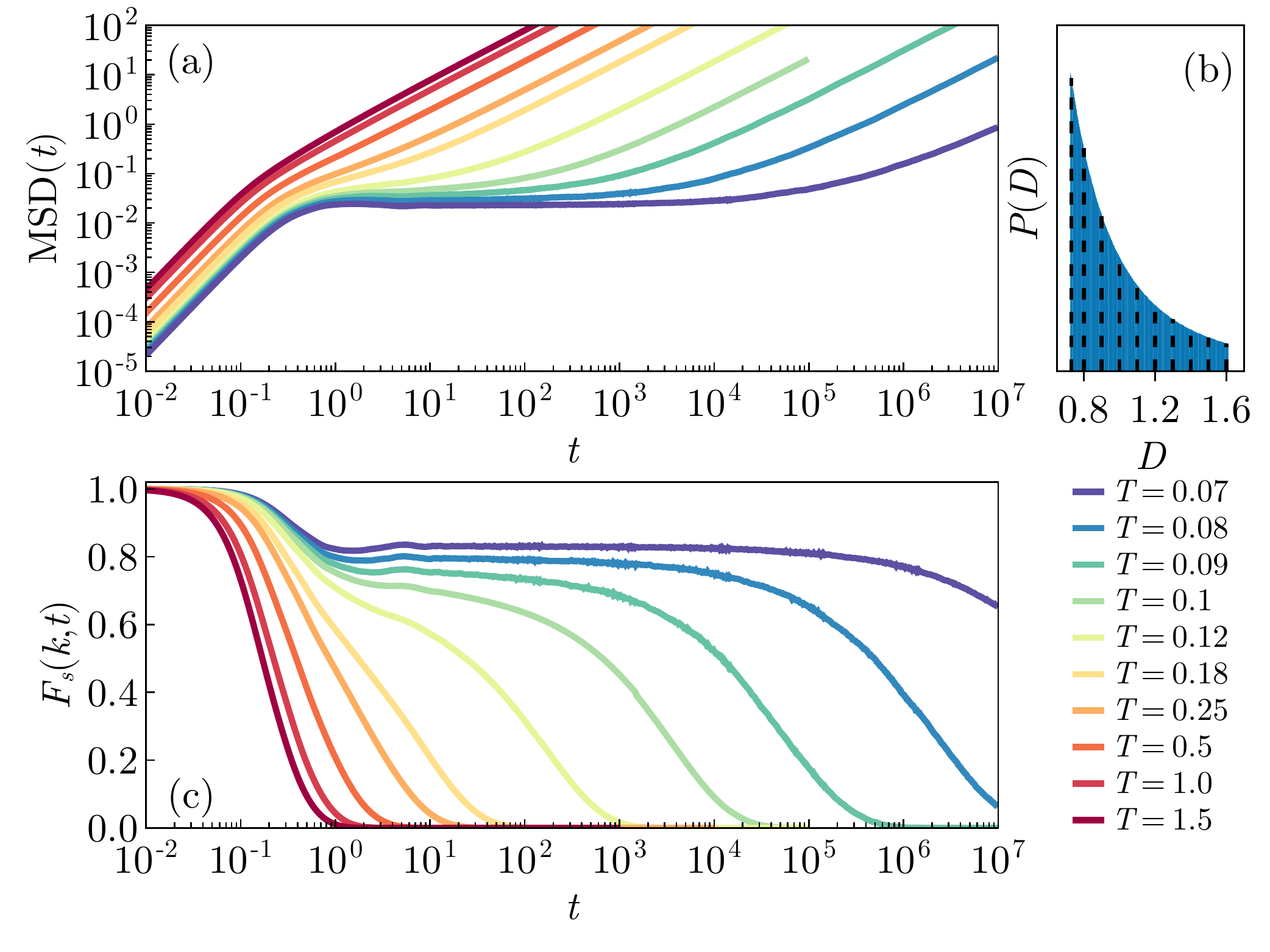}
    \caption{
    Averaged dynamics of the continuously polydisperse mixture. 
    Panels (a) and (c) respectively show the mean squared displacements and self-intermediate scattering functions at wave number $k=7.1$ as a function of time $t$ for different temperatures. Panel (b) shows the particle size distribution of the model, $P(D)=A/D^3$, where $A$ is a normalization constant. The vertical dashed lines indicate the binning procedure performed later in this work.
    }
    \label{fig:methods}
\end{figure}

\section*{Simulation Model}

 We study the three-dimensional model system introduced in Ref.~\cite{ninarello2017models}. Briefly, the particles interact with a pair potential
\begin{equation}
    U(r_{ij}) = \epsilon \left(\frac{D_{ij}}{r_{ij}}\right)^{12} + c_0 + c_2\left(\frac{r_{ij}}{D_{ij}}\right)^{2}+ c_4\left(\frac{r_{ij}}{D_{ij}}\right)^{4}
\end{equation}
within the cutoff $r_{ij}/D_{ij} = 1.25$. Here, $r_{ij}$ is the pair distance between particles $i$ and $j$, and $\epsilon$ is the interaction strength which also serves as our unit of energy. The coefficients $c_0$, $c_2$, and $c_4$ are used to ensure the continuity and differentiability of the potential at the cutoff point. The effective diameter is given by $D_{ij}=\frac{1}{2}(D_i+D_j)(1-\zeta|(D_i-D_j|) $ in which $D_i$ is the diameter of particle $i$, and $\zeta=0.2$ sets the degree of non-additivity to prevent demixing. All particles have mass $m=1$. The particle diameters $D_i$ are sampled from the distribution $P(D) = A/D^3$, with average diameter $\overline{D}=1$, size ratio $D_\mathrm{max}/D_\mathrm{min}=2.219$, and $A$ a normalization constant. The average diameter serves as our unit of length. With this choice of distribution the overall polydispersity is 23\% measured as the ratio of the standard deviation of the distribution and the average diameter. The smallest and largest particles have diameters of 0.73 and 1.62 respectively. This particle size distribution is visualized in Fig.~\ref{fig:methods}(b). 
We set the number density to $\rho = 1$ and the particle number to $N=1200$, which is sufficiently large to avoid finite size effects \cite{scalliet2022thirty}.

For each temperature $T$, we generate at least 120 independent equilibrium initial conditions using SMC. We equilibrate for at least 100$\tau^{\mathrm{SMC}}_\alpha$, where $\tau^{\mathrm{SMC}}_\alpha$ is the $\alpha$-relaxation time of the SMC dynamics given by the time scale at which the self-intermediate scattering function $F_s(k,t)=\sum_j \exp{[i\textbf{k}\cdot(\textbf{r}_j(t)-\textbf{r}_j(0))]}/N$ has decayed to a value of $1/e$ at the wave number $k =|\textbf{k}|=7.1$ corresponding to the peak of the structure factor. In the definition of $F_s(k,t)$, $\textbf{r}_j$ is the position vector of particle $j$. From each generated initial condition, we perform a constant energy Molecular Dynamics (MD) simulation with a time step of $\Delta t=0.01$, in units of $(m \overline D ^2 / \epsilon)^{1/2}$. We simulate $10^9$ MD steps for the lowest temperatures considered in this work ($T=0.07$).  In full agreement with earlier studies, the dynamics exhibit clear caging (i.e.\ a plateau in the MSD) at temperatures below $T_\mathrm{mct} \approx 0.095$, with  $T_\mathrm{mct}$ obtained from a power-law fit of the $\alpha$-relaxation time in Refs.~\cite{scalliet2022thirty, ninarello2017models}. Figure~\ref{fig:methods} shows the mean squared displacements (MSDs) and self-intermediate scattering functions $F_s(k,t)$ obtained from the trajectories, averaged over all particles. Additionally, it shows the particle size distribution.

\section*{Small particles move more}

\begin{figure}[ht]
    \centering
    \includegraphics[width=\linewidth]{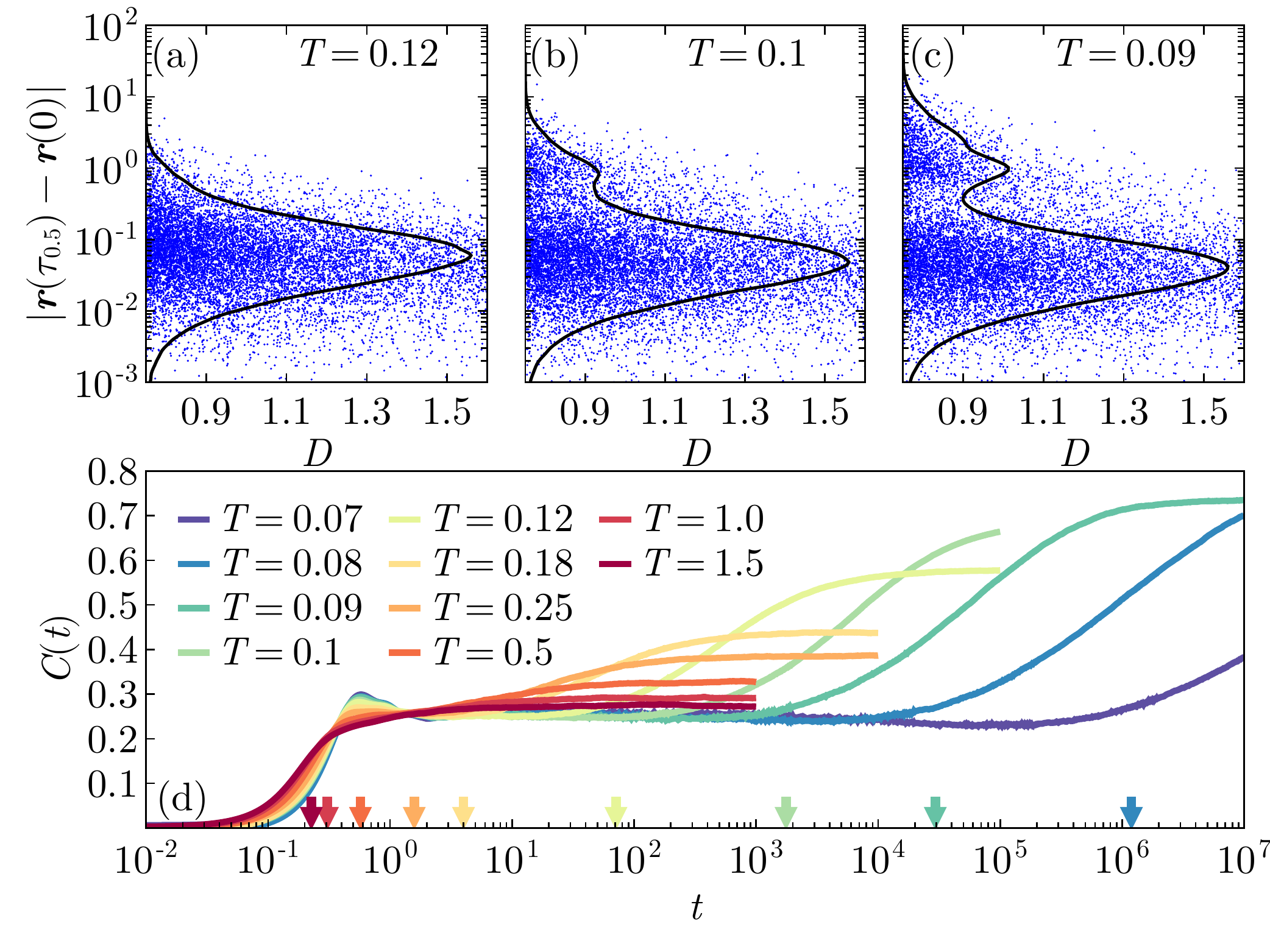}
    \caption{
    \textit{Small particles move more.} Panels (a-c) show absolute displacements scattered against particle size in a sample simulation at the time where the $F_s(k,t)=0.5$ at $k=7.1$. The full black lines display the overall displacement histograms. Panel (d) shows the Pearson correlation coefficient between inverse particle diameter and its absolute displacement as a function of time for different temperatures. For the highest temperatures our simulations are shorter than for lower temperatures. The arrows indicate the $\alpha$-relaxation time.
    }
    \label{fig:pearson}
\end{figure}

As a first measure of the relationship between particle size and mobility, we scatter particle displacements against their size in Fig.~\ref{fig:pearson}(a-c) for different temperatures. At temperatures above the mode-coupling temperature, we see that the absolute displacements are approximately log-normally distributed. However, as the mode-coupling temperature is approached, the distribution of particle displacements broadens, and around $T_\mathrm{mct}\approx0.095$ it transitions into a bimodal distribution. This shows a dynamic separation between mobile and quiescent particles \cite{weysser2010structural, flenner2005relaxation, szamel2006time}. The group of fast-moving particles is disproportionately over-represented by the smaller particles. This separation between small mobile and trapped particles is only observed around timescales associated with cage escape (where $F_s(k,t)$ is between roughly $0.1$ and its plateau value), and is inherently related to dynamical heterogeneities. If we were to probe shorter timescales, virtually all particles would be trapped, whereas at very long timescales, all particles become mobile (see Supplemental Information (SI)). 

\begin{figure}
    \centering
    \includegraphics[width=\linewidth]{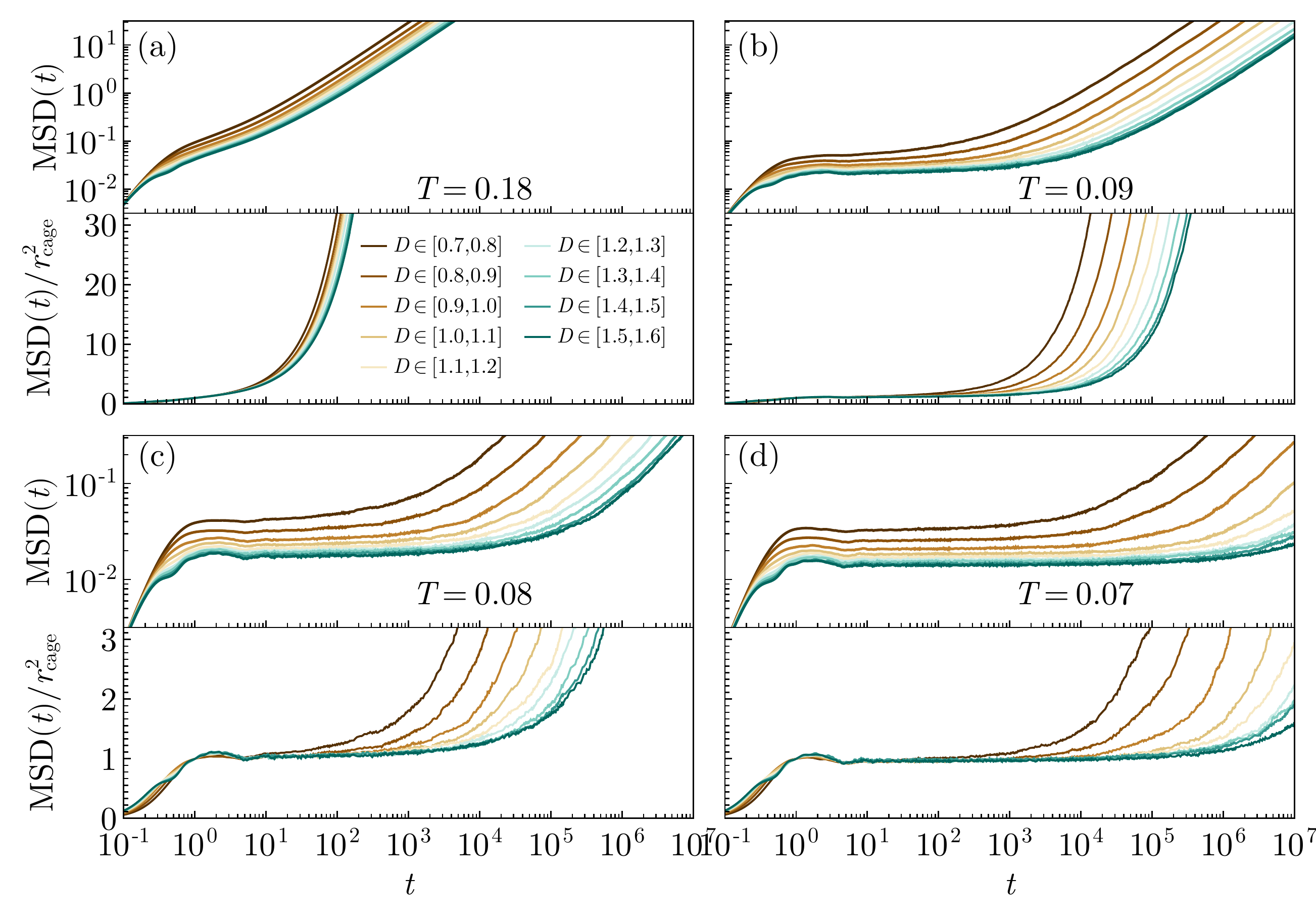}
    \caption{
    \textit{Small particles move first.} We show the MSD for different particle sizes as a function of time for different temperatures around the mode-coupling temperature. The panels (a-d) correspond to $T=0.1, 0.09, 0.08,$ and 0.07 respectively. Each panel is divided in a top part that shows the MSD on double logarithmic axes and a bottom part, that shows the MSD normalized by a measure of the cage size $r_\mathrm{cage}^2 = \mathrm{MSD}(t=1)$ on a linear scale.
    }
    \label{fig:msd}
\end{figure}

\begin{figure}[t]
    \centering
    \includegraphics[width=\linewidth]{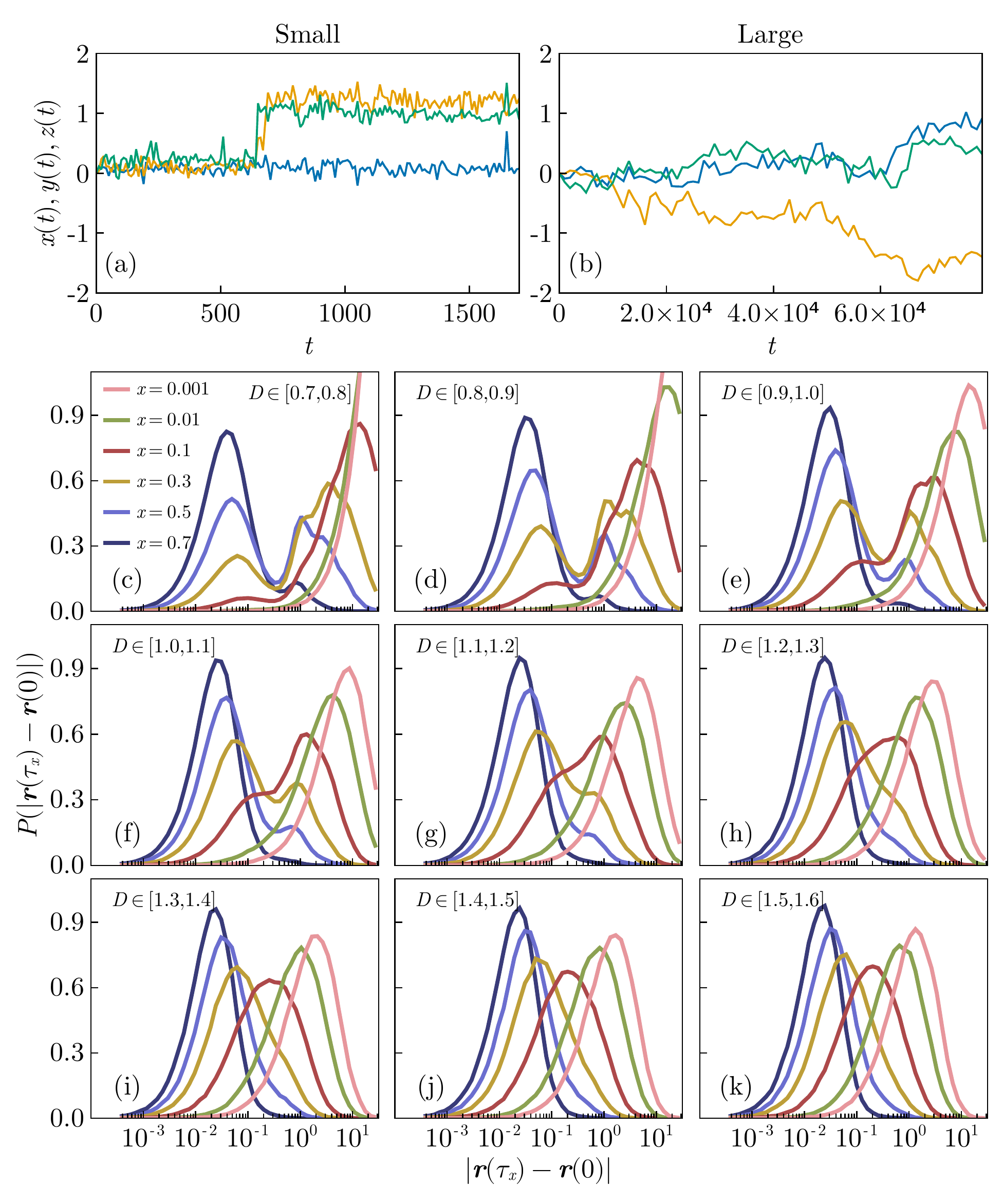}
    \caption{
    \textit{Small particles move differently.} \textbf{Top}: typical trajectory of a small (left) and large (right) particle in the $x$, $y$, and $z$-direction as a function of time for a temperature $T=0.09$. Note that the time span shown is 40 times larger for the large particle compared to the small one. \textbf{Bottom}: particle-size-resolved displacement distributions for different relaxation times $\tau_x$. Each of the nine panels show histograms of absolute displacements for a specific particle size bin. These histograms are computed over the time windows  $\tau_{0.7}\approx6.4\times10^2$, $\tau_{0.5}\approx1.2\times10^4$, $\tau_{0.3}\approx4.6\times10^4$, $\tau_{0.1}\approx1.8\times10^5$, $\tau_{0.01}\approx6.2\times10^5$, and $\tau_{0.001}\approx9.9\times10^5$ at $T=0.09$. The distributions are normalized such that they can be interpreted as probability density functions of $\log_{10}{(|\textbf{r}(\tau_x)-\textbf{r}(0)|)}$.
    }
    \label{fig:hists}
\end{figure}

To quantitatively establish the overall correlation between particle size and mobility, we measure the Pearson correlation between inverse particle size and displacement as a function of time. The Pearson correlation coefficient $\mathrm{cor(x,y)}$ is the standard metric for quantifying correlations between random variables \cite{heumann2016introduction}.
Figure~\ref{fig:pearson}(d) shows the correlation function $C(t) = \mathrm{cor}(1/D, |\textbf{r}(t) - \textbf{r}(0)|)$ between the inverse diameter $1/D$ and absolute particle displacement $|\textbf{r}(t) - \textbf{r}(0)|$ as a function of time. At very short times there exists no correlation between the two. This is because as long as particles have not yet encountered any of their neighbors, their dynamics are independent of their size (in our model the particle mass is independent of particle size). As the particles start encountering neighbors, $C(t)$ increases since cages of smaller particles are slightly larger on average than those of larger particles [as shown later Fig.~\ref{fig:hists} and \ref{fig:msd}(a-d)]. This increase in the correlation between inverse particle size and mobility is virtually temperature independent since the interaction potential is very steep. For sufficiently high temperatures ($T\gg T_\mathrm{mct}$), $C(t)$ finds a steady state, related to the fact that smaller particles have a higher long-time diffusion coefficient than larger particles. The steady-state value is strongly temperature dependent because the dependence of the diffusion coefficient on particle size grows with decreasing temperature \cite{heckendorf2017size, murarka2003diffusion}.

At temperatures close to or below $T_\mathrm{mct}=0.095$, instead of finding a steady state, the correlation peaks around $t\approx 0.6$, which we attribute to caged inertial effects. After the peak, $C(t)$ plateaus around $C(t)=0.25$, as all particles are caged. The plateau value of $C(t)$ encodes the correlation between particle size and its cage size. When particles start to escape their cages, we observe a second growth of $C(t)$, signaling that smaller particles contribute progressively more to the dynamics than larger ones. At time scales beyond around 100 $\alpha$-relaxation times, $C(t)$ finally reaches a steady state \footnote{The fact that it takes at least 100 $\alpha$-relaxation times to ensure that particles reach a fully diffusive state, also suggests that this should be interpreted as a lower limit for the necessary equilibration time. While we do not pursue it in this work, it would be interesting to investigate how this depends on the degree of polydispersity of the mixture.}.

\section*{Small particles move first} In order to investigate the effect of the emergent separation between small, mobile particles and larger, quiescent ones observed above, we divide the particle size distribution into several bins of width $\Delta D=0.1$, as indicated by the black dashed lines in Fig.~\ref{fig:methods}(b). In total, we construct nine bins given by [0.7,0.8], [0.8,0.9], $\ldots$, [1.5,1.6]. Since the distribution is asymmetric, the bins contain progressively fewer particles. Due to the functional form of the size distribution, each bin contains roughly the same total particle volume fraction. The high number of independent simulations we perform for each temperature ensures that we have sufficient statistics. In the remainder of this section, we focus on the particle-size resolved MSD to support our observations. We have verified that other proxies of mobility such as the self-intermediate scattering functions and bond-breaking correlation functions are also in line with our findings (See Supplemental Information (SI)).

In Fig.~\ref{fig:msd} we show the size-resolved MSDs as a function of time for different temperatures. Each of these four panels depicts the MSDs both on a logarithmic grid (top) and on a linear grid, the latter normalized by an approximate measure of the squared cage size $r_\mathrm{cage}^2 = \mathrm{MSD}(t=1)$ (bottom). This length scale $r_\mathrm{cage}$ serves only for illustrative purposes and our results do not depend on its precise definition.
It can be seen that, upon cooling across the mode-coupling temperature, the difference in mobility between particles of different sizes grows continuously, culminating in differences of orders of magnitude between small and large particles at the lowest temperatures studied (see SI for a plot of this ratio). Our data shows that not only  the small particles are more mobile, but they also escape their cages first. 
This observation is most apparent in the bottom panels of Fig.~\ref{fig:msd}(c-d), which unambiguously show that the particles of size $D\in [0.7,0.8]$ start to escape their cages decades in time before the larger particles start to do so. The MSD curves also confirm the earlier observation that small particles move significantly more than larger ones. The finding that small particles move first sheds new light on the recent discovery that the dynamics in this model system are highly dominated by dynamic facilitation \cite{scalliet2022thirty, garrahan2002geometrical, chandler2010dynamics}.
Indeed, since the relaxation of smaller particles clearly precedes that of larger ones, our results suggest that the former facilitates the cage escape of the latter.

\section*{Small particles move differently}
We now focus on the differences between how small and large particles move through the supercooled liquid at temperatures near the dynamic crossover temperature $T_\mathrm{mct}$. To do so, we compute histograms of particle displacements over different time periods $\tau_x$, defined as $F_s(k, \tau_x)=x$, for the different particle size bins. With this definition, $\tau_{1/e}$ corresponds to the standard $\alpha$-relaxation time. We focus on the wave number $k=7.1$, i.e.\ the main peak of the static structure factor.
The histograms are shown in Fig.~\ref{fig:hists}(c-k) for $T=0.09$, which is just above the crossover temperature. In Fig.~\ref{fig:hists}(a-b) we also show a typical trajectory of a small and a large particle respectively. The histograms for different temperatures are presented in the SI.

The displacement distributions in Fig.~\ref{fig:hists}(c-k) show again the appearance of the distinct double peak in the displacement distributions of small particles at intermediate timescales. It indicates that the population of small particles separates into a quiescent and a mobile subpopulation, respectively acquiring displacements on the order of $0.1D$ (for caged particles) and on the order of $D$ (for the particles that have cage-hopped). As time progresses, the small particles `hop' out of the quiescent population into the mobile one as they undergo cage-escape. At long timescales, such as $\tau_{0.01}$, almost all small particles have undergone several cage-hopping events, leading to a distribution with a single peak at distances much greater than a particle diameter. The hopping-like motion can also be clearly observed in the example trajectory of the small particle in Fig.~\ref{fig:hists}(a).

Shifting our attention to the larger particles, (Fig.~\ref{fig:hists}(g-k)), one might expect that a similar hopping regime would occur, but at later times. This, however, is not the case. The large particle displacement distribution does not split into two peaks at any time, but gradually shifts to larger displacements. The distributions show only a very mild increase in heterogeneity during this process. These observations imply that the dominant relaxation pathway of small particles, cage hopping, is subdominant for larger particles. Instead, the larger particles experience something more akin to standard---albeit very viscous---diffusion. We expect that the observed differences between large and small particles become even more extreme at temperatures below the mode-coupling temperature.

The difference in relaxation mechanisms between small and large particles can be rationalized with the limiting picture of a very large particle in a background supercooled liquid made of comparably tiny particles. The large particle moves through the liquid not by displacing the particles in its way or `hopping' from cage to cage, but essentially by being displaced by the reorganizations of smaller, more mobile particles. In this sense, the background supercooled liquid is dynamically facilitating the motion by the same process that underlies Brownian motion. This picture is supported by the example trajectory in Fig.~\ref{fig:hists}(b), which shows random-walk-like motion, rather than the hopping-like motion of small particles.

A limiting picture can also be drawn in the limit of a very small tagged particle in an environment of much larger particles. Because it is so small, the tagged particle can move through the liquid without significantly displacing the surrounding particles, like a tracer in a non-percolating random Lorentz gas \cite{biroli2021mean}. It can thus undergo non-cooperative hopping motion, and as such it is natural that small particles initiate structural relaxations.

Of course, the degree of polydispersity in the current system is not sufficient (and the temperature too high) to see complete separation between particles that undergo the two processes. In particular, Fig.~\ref{fig:hists} shows that while large particles predominantly undergo diffusive-like motion, at intermediate timescales the displacement distribution broadens, indicating an increased dynamic heterogeneity that suggests some degree of hopping-like motion. Similarly, while small particles mainly move by undergoing cage escape, the left peak in their displacement distribution, which is associated with caged dynamics, does shift to the right as time progresses. This indicates that they too experience standard diffusive motion to some extent.

In the less polydisperse binary Kob-Andersen mixture, a similar distinction between hopping and diffusion 
 of small and large particles is found above the mode-coupling temperature \cite{kob1995testing,flenner2005relaxation}. In that system, below some temperature $T>T_\mathrm{mct}$, hopping becomes the dominant relaxation mechanism also of the large particle species. In the current model system, however, the distinction remains present well below $T_\mathrm{mct}$, at least down to $T=0.08$ (see SI). Our simulations are not sufficiently long to conclusively establish whether there exists an even lower temperature below which hopping becomes dominant for all particle sizes. 

\begin{figure}
    \centering
    \includegraphics[width=\linewidth]{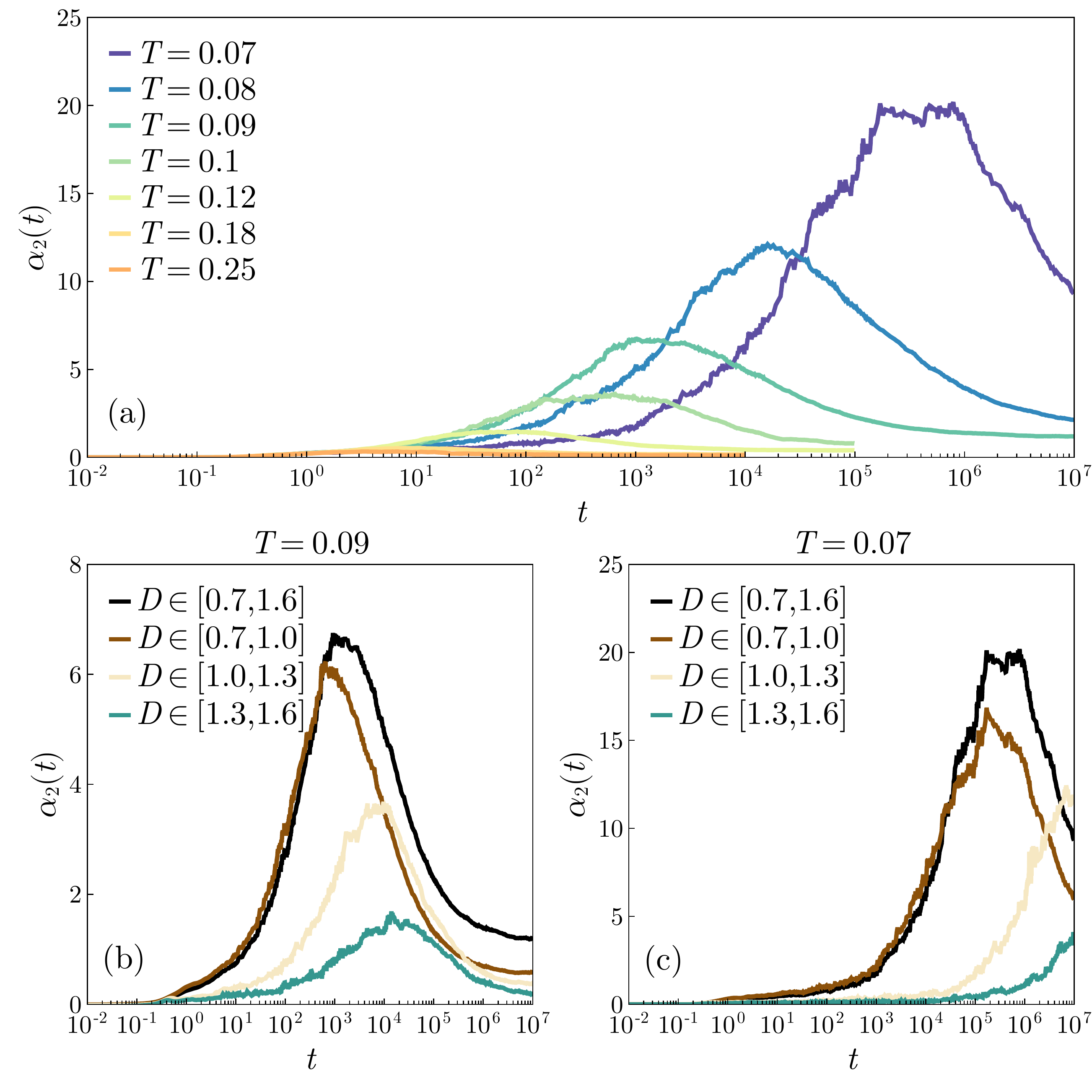}
    \caption{Quantifying dynamic heterogeneity. Panel (a) shows the total non-Gaussian parameter $\alpha_2(t)$ as a function of $t$ for varying temperatures $T$. Panels (b) and (c) show non-Gaussian parameters for specific particle sizes at temperatures $T=0.09$ and $T= 0.07$, respectively.}
    \label{fig:alpha_2}
\end{figure}

In order to quantify the degree of temporal heterogeneity, we compute the total, species agnostic non-Gaussian parameter \cite{rahman1964correlations}
\begin{equation}
    \alpha_2(t) = \frac{3}{5} \frac{\left<|\textbf{r}(t) - \textbf{r}(0)|^4\right>}{\left<|\textbf{r}(t) - \textbf{r}(0)|^2\right>^2} - 1.
\end{equation} 
The results are shown in Fig.~\ref{fig:alpha_2}(a). We find that the non-Gaussian parameter displays a peak which grows as one descends into the supercooled regime. This peak corresponds to the timescale at which the degree of dynamic heterogeneity is the largest, and is typical for the supercooled regime.  At long times, we find that $\alpha_2(t)$ converges to a constant, which indicates that not all particles have the same diffusivity. This will be further quantified in the next section. 

To further illustrate the greater dynamic heterogeneity of the smallest particles compared to the larger ones, we display in Fig.~\ref{fig:alpha_2}(b)-(c) the non-Gaussian parameters for specific particle sizes. In order to improve statistics, we use broader bins than in the rest of the work. Our results show that below $T_{\text{mct}}$, the degree of heterogeneity of the smallest particles is substantially greater than that of the largest particles. Furthermore we again observe the decoupling in relaxation times as the peak of the non-Gaussian parameter, which is proportional to the structural relaxation time $\tau_{\alpha}$, occurs decades earlier for the smallest size particles compared to the larger ones. This confirms once more that small particles move first. Lastly, Fig.~\ref{fig:alpha_2}(b)-(c) shows that, at early times, the diameter-agnostic non-Gaussian parameter (black line) is completely dominated by that of the smallest particles. This is consistent with the data shown in Fig.\ref{fig:hists}(c)-(k), from where it is clear that on timescales shorter than the $\alpha$-relaxation time, the small particles have started cage-hopping, whereas the intermediate and larger ones are still mostly trapped. 
Hence, 
diameter-agnostic quantities which mask such species-specific dynamical features may not paint a complete physical picture of the dynamics.

\section*{The dynamic crossover}

\begin{figure*}
    \centering
    \includegraphics[width=\linewidth]{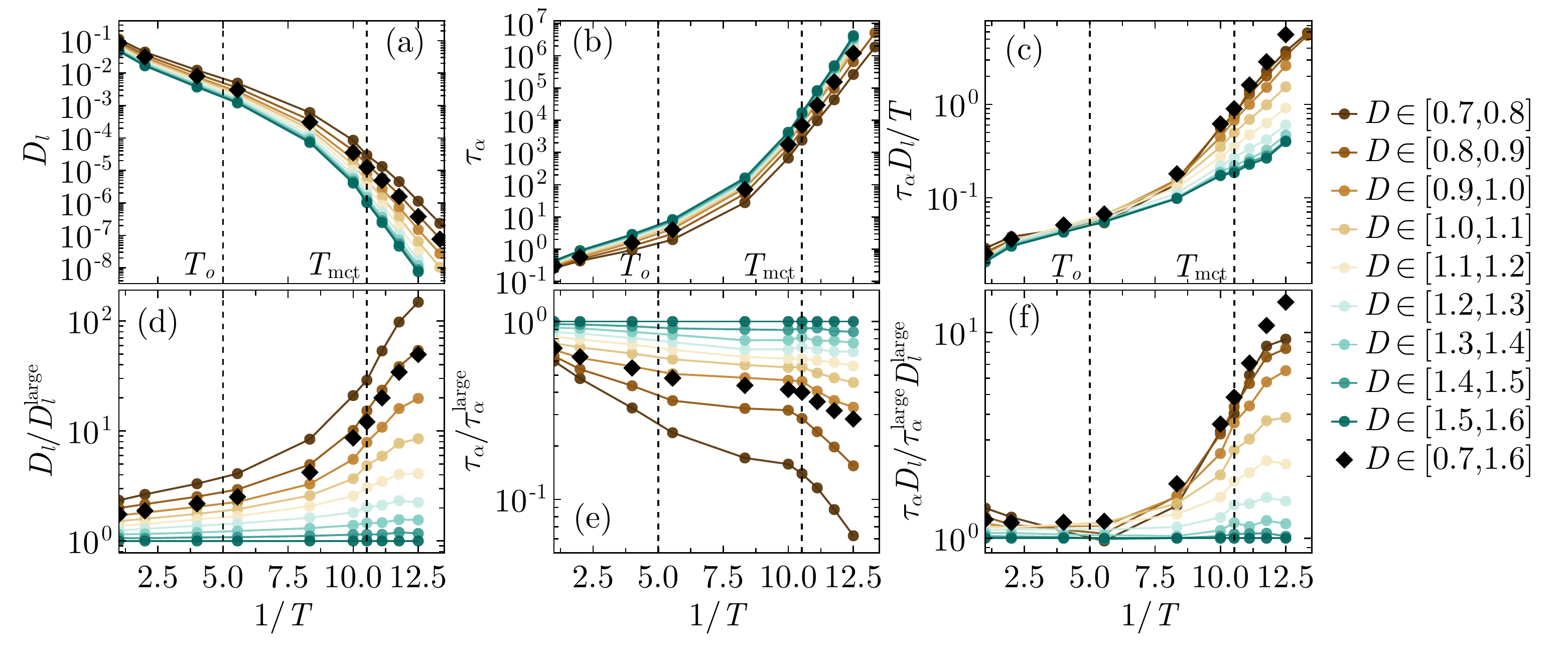}
    \caption{
    Temperature dependence of (a) the diffusion constant $D_l$, (b) the $\alpha$-relaxation time $\tau_\alpha$, and (c) the product $D_l\tau_\alpha/T$ for different species. The latter is a measure of the Stokes-Einstein violation. Panels (d-f) show the same data, but normalized by the values for the largest species. Black diamonds are size-agnostic quantities. The vertical dashed lines indicate the onset temperature $T_o=0.2$ and the mode-coupling temperature $T_\mathrm{mct}=0.095$ as determined in Ref.~\cite{scalliet2022thirty}. 
    }
    \label{fig:Dta}
\end{figure*}

In order to more quantitatively establish the effect of polydispersity on the power-law-to-Arrhenius crossover of the relaxation time, we display in Figs.~\ref{fig:Dta}(a) and \ref{fig:Dta}(b) the temperature dependence of the long-time diffusion coefficient $D_l$ and the $\alpha$-relaxation time $\tau_\alpha = \tau_{1/e}$. For convenience, we also indicate the onset temperature $T_o$ of the power-law regime, and the mode-coupling temperature that marks the crossover to Arrhenius scaling. We clearly observe both the power law  $\tau_\alpha, D_l \propto |T-T_\mathrm{mct}|^{\pm\gamma}$ and Arrhenius regime $\tau_\alpha, D_l \propto \exp({\pm E_{a}/T})$ in the temperature scaling of both the diffusion constant and the relaxation time. Here, $\gamma$ is the mode-coupling exponent and $E_a$ is a characteristic activation energy. The crossover we observe between the two regimes agrees well with the earlier reported value of $T_\mathrm{mct}=0.095$ \cite{ninarello2017models, scalliet2022thirty}. In \ref{fig:Dta}(c), we show $\tau_\alpha D_l/T$, which is a measure for the degree of the breakdown of the Stokes-Einstein violation \cite{shi2013relaxation, zaccarelli2015polydispersity, sengupta2014distribution}. It is clear that the breakdown is the strongest among the smallest particles, which can be explained by their larger dynamic heterogeneity.

In Figs.~\ref{fig:Dta}(d-e), we respectively show the same data as in Figs.~\ref{fig:Dta}(a-c) normalized by those of the largest particle species. This allows us to easily distinguish differences in temperature scaling between the different particle sizes. Clearly, if the MCT point $T_\mathrm{mct}$ and associated exponent $\gamma$ were size independent, the resulting curves would approach some constant. The same is true for the activation energies $E_a$ in the Arrhenius regime. Focusing on the diffusion constants in the power-law regime (Fig.~\ref{fig:Dta}(d)), we do not observe such a plateau, indicating that at least one of the associated $\gamma$ or $T_\mathrm{mct}$ for the diffusion constant must be size dependent. As we have only 3 data points in the power-law regime, we cannot reliably resolve this size dependency further. 
This qualitatively corroborates the findings of Zaccarelli and coworkers who show similar results for a system of hard particles with around 10\% polydispersity \cite{zaccarelli2015polydispersity}. 

The relative relaxation times do however show a plateau as $T_\mathrm{mct}$ is approached. This indicates that the exponent and critical point in the mode-coupling scaling law $\tau_\alpha \propto |T-T_\mathrm{mct}|^{\gamma}$ are (nearly) independent of particle size, as predicted by the theory \cite{gotze2009complex}. Since the same is not true for the scaling of the diffusion coefficient, as we discussed above, we can thus attribute the latter to the breakdown of the Stokes-Einstein relation, which mode-coupling theory does not account for. From the Arrhenius regime $T<T_\mathrm{mct}$, both for $\tau_\alpha$ and $D_l$, we can also infer that the activation energy is size-dependent. 

\section*{Conclusion} Our results establish that strongly particle-size-dependent effects emerge in the dynamics of highly polydisperse supercooled liquids near the mode-coupling temperature. We find that structural relaxation is initiated by the cage escape of small particles and that the movement of the large particles is mainly facilitated by that of smaller particles. The dominant relaxation mechanism of small particles is cage hopping, whereas that of large particles is more akin to standard diffusion. By studying the long-time dynamics, we also find that the degree to which the Stokes-Einstein relation breaks down, as well as the distribution of activation energy barriers in the supercooled regime, are dependent on the particle size. While these striking differences are inherently masked in the particle-averaged dynamics, they are in fact crucial to fully understanding the observed structural relaxation. 

In view of the significant difference between timescales governing the dynamics of small and large particles, it is natural to ask to what extent observations of the power-law tails and wings in \textit{e.g.}\ waiting time distributions and relaxation spectra are unique to the imposed size polydispersity \cite{zaccarelli2015polydispersity}. 
Indeed, it is known that introducing additional relaxation channels in the form of orientational degrees of freedom or dynamic particle sizes has a large effect on dynamic facilitation and the glass transition \cite{mishra2014dynamical, berthier2019efficient}. Given that our results here point to the fact that differently sized particles have different relaxation channels, it is not unthinkable that high size polydispersity exerts similar effects. We believe this avenue of research should be explored further to fully revel in the new vistas afforded by enhanced sampling techniques in the deeply supercooled regime.

\section*{Acknowledgments} We thank Vincent Debets and Daniele Coslovich for their insightful comments and critical reading of the manuscript. We acknowledge the Dutch Research Council (NWO) for financial support through a Vidi grant (IP, CCLL, and LMCJ) and START-UP grant (LMCJ). 

\bibliography{apssamp}

\end{document}